\documentclass[11pt,a4paper]{article}
\usepackage[utf8]{inputenc}
\usepackage[T1]{fontenc}
\usepackage{lmodern}
\usepackage{microtype}

\usepackage{setspace}
\usepackage{amsmath}
\usepackage{amssymb}
\usepackage{amsthm}

\usepackage{algorithm}
\usepackage{algpseudocode}

\usepackage{graphicx}

\usepackage{booktabs}
\usepackage{multirow}

\usepackage{url}

\usepackage{natbib}

\usepackage{threeparttable}

\usepackage{footnote}
\makesavenoteenv{tabular}
\makesavenoteenv{table}

\usepackage[hidelinks]{hyperref}
\hypersetup{
    pdftitle={Creative Fatigue Screening with Path Signatures},
    pdfauthor={Charles Shaw}
}

\title{\textbf{Creative Fatigue Screening with Path Signatures}}
\author{Charles Shaw\\WPP Data Science\\\small Research initiated at T\&P}
\date{September 2026}

\begin{document}
\maketitle

\begin{abstract}
Creative fatigue is an operational concern in advertising, but a decline in observed performance does not by itself identify fatigue: delivery, targeting, auction conditions and reporting can produce similar patterns. We study a screening rule for prioritising creative review from daily impressions and clicks. The method smooths click-through rate (CTR), standardises it against an initial calibration period, compares truncated signatures of adjacent time--CTR paths, and requires a separate directional mean-drop gate. This separation matters because ordinary path signatures are translation invariant: calibration scaling preserves the amplitude of within-window movements, whereas the gate supplies the absolute deterioration criterion.

We evaluate the hybrid rule on 264 synthetic daily trajectories: 72 with known fatigue-like onsets and 192 no-event or nuisance controls. At the headline operating point, it detects 68 of 72 events (94\%), with a six-day median delay among hits, and alerts on 60 of 192 controls (31\%). A hand-engineered adjacent-window feature comparator is effectively tied at 94\% event recovery and 33\% false alerts. Ablations show that the mean-drop gate carries most event recovery; the signature threshold reduces nuisance burden in this data-generating process. A full-series offline reference performs better but uses future observations. These results support further evaluation of a portfolio screening workflow, not causal fatigue attribution, production validation, or automatic creative retirement.
\end{abstract}


\section{Introduction}
\label{sec:introduction}

Creative assets can lose effectiveness after repeated exposure, but the timing and form of wear-out vary across messages, audiences and media contexts \citep{pechmann1988advertising,bass2007wearout}. The operational decision is therefore not simply whether wear-out exists. It is which assets merit review, and when, across a portfolio of repeated decisions.

Observed advertising performance makes that decision difficult. Daily CTR is a noisy ratio whose variance depends on impressions. Delivery and reporting may change over time. Audience, placement, auction and campaign composition may also move. A persistent decline is compatible with creative fatigue, but it is not unique to it. Screening must therefore be separated from causal diagnosis.

Changepoint methods provide established tools for detecting distributional change \citep{aminikhanghahi2017survey,truong2020selective}. Path signatures offer a complementary representation of ordered trajectories: truncated iterated integrals encode displacement and higher-order interactions without prescribing one response curve \citep{chevyrev2016primer,fermanian2021embedding}. This paper asks a narrow applied question: can signature-based adjacent-window features contribute to a practical screen for fatigue-like deterioration?

The contribution is a \emph{hybrid research detector}, not a signature-only fatigue model. It combines:
\begin{enumerate}
    \item a calibration-scaled path-signature distance for unusual changes in ordered local trajectory;
    \item a directional adjacent-window mean-drop gate that restricts alerts to deterioration;
    \item a past-only threshold and first-alert rule suitable for retrospective evaluation of online use; and
    \item a synthetic benchmark that reports event recovery, alert delay and nuisance-alert burden against simpler comparators.
\end{enumerate}

The distinction between the first two components is substantive. An ordinary signature depends on path increments and is invariant to translation. Moving an otherwise identical CTR path down by a constant does not change its signature. Calibration scaling avoids the additional information loss caused by independently rescaling every window, but absolute mean-level decline still enters through the separate gate.

\subsection{Evidence boundary}
\label{sec:evidence-boundary}

The evidence is synthetic. It tests code behaviour under specified data-generating processes (DGPs); it does not validate performance on client campaigns. The trajectories are complete daily histories. A low-delivery scenario reduces impressions but does not create missing calendar days, so the experiment does not establish robustness to genuine observation gaps or irregular sampling. The alert threshold is heuristic rather than a significance test or controlled false-alarm procedure. Finally, a screen identifies candidates for contextual review: it does not establish that creative fatigue caused the observed change, quantify causal financial impact, or automate refresh decisions.

\section{Related Work and Positioning}
\label{sec:related-work}

Advertising wear-in and wear-out have long been studied as dynamic response phenomena \citep{pechmann1988advertising,naik1998planning,bass2007wearout}. Research on online advertising also shows that response depends on creative and impression history \citep{braun2013online}. These traditions motivate fatigue monitoring, but they do not make a change in aggregate CTR causally attributable to the creative.

Statistical changepoint methods include cumulative-sum procedures, moving-window statistics, likelihood-based rules, state-space models and offline segmentation \citep{aminikhanghahi2017survey,truong2020selective,adams2007bayesian,killick2012optimal}. They address important parts of the problem. The research question here is not whether those methods exist, but whether path features add useful ordered-trajectory information within an operational screen.

For a bounded-variation path \(X:[0,1]\rightarrow\mathbb{R}^{m}\), the signature is the sequence of iterated integrals
\begin{equation}
S(X)=\left(1,\int dX,\int_{0<u_1<u_2<1}dX_{u_1}\otimes dX_{u_2},\ldots\right).
\end{equation}
The full signature determines a path only up to tree-like equivalence under appropriate qualifications \citep{hambly2010uniqueness}; a finite truncation has no corresponding uniqueness guarantee. In machine learning, truncated signatures are finite feature maps for ordered data \citep{chevyrev2016primer,fermanian2021embedding}. This paper uses ordinary signatures of ordered piecewise-linear paths. It does not implement or evaluate a jump-flow extension.

\section{Hybrid Screening Method}
\label{sec:method}

\subsection{Observed series and smoothing}

For day \(t\), let \(I_t\) be impressions and \(C_t\) clicks, with observed
\[
CTR_t=C_t/I_t.
\]
The benchmark assumes \(C_t\mid I_t,p_t\sim\mathrm{Binomial}(I_t,p_t)\). To reduce exposure-dependent sampling noise, we form a Beta--Binomial posterior mean. Its prior mean is the impression-weighted CTR in the first \(T_C=45\) days and its prior strength is 1,000:
\begin{equation}
\widehat p_t=\frac{C_t+\alpha}{I_t+\alpha+\beta}.
\end{equation}
The detector uses a trailing seven-day mean \(y_t\) of \(\widehat p_t\). This is a pragmatic smoother; it induces serial dependence and is not a fitted latent-state model.

Let \(c\) and \(s\) be the mean and sample standard deviation of \(y_t\) over the calibration period. Define
\begin{equation}
z_t=\frac{y_t-c}{\max(s,\varepsilon)},\qquad \varepsilon=10^{-6}.
\end{equation}
The same calibration transformation is applied to all subsequent windows.

\subsection{Adjacent-window signature distance}

At candidate boundary \(b\), take adjacent windows of \(w=14\) observations:
\[
L_b=(z_{b-w},\ldots,z_{b-1}),\qquad
R_b=(z_b,\ldots,z_{b+w-1}).
\]
Each window is embedded as a piecewise-linear path \(X(u_i)=(u_i,z_i)\), where \(u_i\) is equally spaced on \([0,1]\). Let \(S^{\leq d}(X)\) be its signature truncated at depth \(d=3\). The trajectory-change statistic is
\begin{equation}
D_b=\left\|S^{\leq d}(X_{L_b})-S^{\leq d}(X_{R_b})\right\|_2.
\label{eq:signature-distance}
\end{equation}
An exposure-augmented variant adds calibration-standardised log impressions as a third path coordinate.

The label \emph{calibration-scaled signature} is deliberate. Because the signature is computed from increments, it does not encode an additive level offset. At depth one the time increments cancel and \(D_b\) compares the endpoint displacement of the two CTR paths. Higher levels add ordered interactions between time and CTR, but no finite truncation is a complete representation.

\subsection{Past-only threshold and directional gate}

For each eligible boundary, the reference set contains only earlier signature distances. With at least \(H=25\) earlier distances, define
\begin{equation}
h_b=\overline D_{<b}+k\,s(D_{<b}),
\qquad k=1.5.
\label{eq:threshold}
\end{equation}
The directional statistic is
\begin{equation}
G_b=\frac{\overline y_{L_b}-\overline y_{R_b}}{\max(s,\varepsilon)}.
\label{eq:gate}
\end{equation}
The hybrid rule requires \(D_b>h_b\) and \(G_b>\gamma\), with \(\gamma=2\). It returns the first qualifying alert. The boundary \(b\) is the first observation in the right window, but the actionable issue day is \(b+w-1\), when that window has been observed.

\begin{algorithm}[htp!]
\caption{Past-calibrated hybrid signature screen}
\label{alg:hybrid}
\begin{algorithmic}[1]
\Require daily impressions \(I_t\), clicks \(C_t\); \(T_C,w,d,H,k,\gamma\)
\State form posterior CTR and its trailing seven-day mean \(y_t\)
\State estimate \(c,s\) from days \(1,\ldots,T_C\); calculate \(z_t\)
\For{each adjacent-window boundary \(b\)}
    \State compute \(D_b\) from Equation \ref{eq:signature-distance}
    \State compute \(G_b\) from Equation \ref{eq:gate}
\EndFor
\For{each boundary \(b\) in time order}
    \If{\(b\geq T_C\), at least \(H\) earlier distances exist, \(D_b>h_b\), and \(G_b>\gamma\)}
        \State \Return first actionable alert at \(b+w-1\)
    \EndIf
\EndFor
\State \Return no alert
\end{algorithmic}
\end{algorithm}

The threshold is an empirical operating rule. Overlapping windows make the distance history dependent, and repeated testing across time and creatives is not calibrated here to a family-wise error rate or false-discovery rate.

\subsection{Research detector versus current software}

The research detector above is not the behaviour of the current CreativeDynamics library or application. The current library independently min--max scales adjacent windows, compares shape-signature distances, uses a full-series retrospective threshold, classifies resulting segments, and generates reports and impact summaries. It does not use calibration scaling, the past-only threshold, the mean-drop gate or the exposure coordinate evaluated here. The research detector should therefore be treated as a candidate specification pending implementation and external validation.

\subsection{Computational order}

For path dimension \(m\), depth \(d\) and window length \(w\), a direct truncated-signature calculation retains \(\sum_{j=1}^{d}m^j\) coordinates. Its approximate cost per window is
\[
\mathcal O\!\left(w\sum_{j=1}^{d}m^j\right)
\]
and across \(N\) candidate boundaries it is
\[
\mathcal O\!\left(Nw\sum_{j=1}^{d}m^j\right).
\]
The detector is linear in \(N\) only when \(w,m,d\) are fixed; feature dimension grows exponentially with depth.

\section{Synthetic Evaluation Design}
\label{sec:evaluation}

\subsection{Data-generating processes}

The deterministic benchmark uses seed 20260425, 24 replicates of 11 scenarios, 180 daily observations, and a reference onset at day 90. Three scenarios contain known fatigue-like events: an abrupt level drop, a gradual decline and a volatile decline. Eight controls represent stability or nuisance mechanisms: impression shift, seasonality shift, reduced-delivery intervals, targeting-mix shift, auction pressure, attribution volatility and creative rotation.

Impressions combine lognormal series-level scale, weekly variation, lognormal daily variation and Poisson sampling. Clicks are binomial conditional on impressions and latent CTR, except that the attribution-volatility control alters reported clicks after sampling. Event scenarios start at latent CTR 2.4\%; most controls start at 1.8\%. This baseline difference is a design limitation: it makes the benchmark useful for software behaviour and relative operating points, but not a population-calibrated estimate of performance.

\begin{table}[htp!]
\centering
\caption{Synthetic scenario diagnostics around the injected reference day}
\label{tab:scenario_diagnostics}
\scriptsize
\setlength{\tabcolsep}{2pt}
\resizebox{\textwidth}{!}{%
\begin{tabular}{p{3.0cm}lccc}
\toprule
\textbf{Scenario} & \textbf{Type} & \textbf{Latent CTR pre/post} & \textbf{Observed CTR pre/post} & \textbf{Impr. pre/post} \\
\midrule
Stable & No event & 1.80/1.80\% & 1.81/1.80\% & 7040/7076 \\
Impression shift & No event & 1.80/1.80\% & 1.80/1.80\% & 8466/3842 \\
Seasonality shift & No event & 1.80/1.80\% & 1.80/1.80\% & 7704/7704 \\
Delivery gaps & No event & 1.80/1.80\% & 1.81/1.79\% & 7349/7418 \\
Targeting mix & No event & 1.80/1.45\% & 1.80/1.45\% & 8411/8550 \\
Auction pressure & No event & 1.80/1.80\% & 1.80/1.81\% & 7591/5460 \\
Attribution volatility & No event & 1.80/1.80\% & 1.76/1.84\% & 8355/8448 \\
Creative rotation & No event & 1.81/1.79\% & 1.81/1.80\% & 7956/8383 \\
Abrupt drop & Event & 2.40/1.50\% & 2.39/1.50\% & 7390/7528 \\
Gradual decline & Event & 2.40/1.66\% & 2.39/1.65\% & 8201/8328 \\
Volatile decline & Event & 2.40/1.50\% & 2.40/1.50\% & 7561/7611 \\
\bottomrule
\end{tabular}
}
\end{table}

The \emph{delivery gaps} label in Table \ref{tab:scenario_diagnostics} denotes intervals of sharply reduced impressions. All 180 calendar days remain observed and impressions are bounded below at 200. Genuine missingness, irregular observation times and zero-delivery periods remain unevaluated.

\subsection{Comparators and scoring}

All online comparators receive the same smoothed CTR series and are scored at their actionable alert day. They comprise:
\begin{itemize}
    \item shape-only signatures with independent per-window min--max scaling;
    \item a four-feature adjacent-window distance using mean, slope, standard deviation and drawdown;
    \item a Gaussian-kernel maximum mean discrepancy window statistic;
    \item GLR-style and MOSUM-style adjacent-window mean-drop statistics; and
    \item a one-sided CUSUM drop rule.
\end{itemize}
The labels describe implemented statistics rather than exhaustive method families. An offline single-mean-shift reference uses the full series and is not an online comparator.

An event hit is the first alert in \([90,135]\). An alert outside that interval on an event series is invalid. Any alert on a control is a false alert. Delay is reported only among detected events. This conditioning must be read with the hit rate. The primary results use fixed operating settings. A secondary split selects settings on the first 12 replicates of every scenario and freezes them for the remaining 12; because train and held-out sets share scenario families, this is a leakage check rather than external validation.

\subsection{Operational proxy}

For event scenarios only, the code defines daily benchmark-relative shortfall
\[
q_t=I_t\max(0,\bar p_C-\widehat p_t),
\]
where \(\bar p_C\) is the mean posterior CTR during calibration. Summing \(q_t\) before or after an alert gives a click-equivalent opportunity proxy for synthetic comparisons. It is not observed lost clicks, causal value, spend wastage or return on investment. No monetary claim is made.

\section{Results}
\label{sec:results}

\subsection{Headline operating point}

\begin{table}[htp!]
\centering
\caption{Injected changepoint benchmark: 72 event series and 192 controls}
\label{tab:injected_benchmark}
\footnotesize
\setlength{\tabcolsep}{3pt}
\resizebox{\textwidth}{!}{%
\begin{tabular}{lcccc}
\toprule
\textbf{Method} & \textbf{Event hit} & \textbf{Median delay} & \textbf{Invalid alert} & \textbf{Control alert} \\
\midrule
Signature shape-only & 26\% & 15 & 72\% & 99\% \\
Cal.-scaled sig. retros. & 29\% & 6 & 71\% & 25\% \\
Cal.-scaled sig. & 94\% & 6 & 6\% & 31\% \\
Cal.+exposure sig. & 96\% & 6 & 4\% & 32\% \\
Window features & 94\% & 6 & 4\% & 33\% \\
Kernel window & 93\% & 11 & 7\% & 47\% \\
GLR mean-drop & 90\% & 6 & 10\% & 49\% \\
MOSUM mean-drop & 50\% & 6 & 50\% & 93\% \\
CUSUM drop & 78\% & 5 & 22\% & 57\% \\
Offline full-series ref. & 100\% & 4 & 0\% & 12\% \\
\bottomrule
\end{tabular}
}
\end{table}

The hybrid calibration-scaled signature rule detects 68 of 72 events (94.44\%) and alerts on 60 of 192 controls (31.25\%); Wilson 95\% intervals are approximately 86.6--97.8\% and 25.1--38.1\%, respectively. Median delay is six days among hits. The four-feature comparator also detects 68 of 72 events and alerts on 63 of 192 controls (32.81\%). These differences are too small, under a single seeded simulation design, to support superiority of signatures.

The exposure-augmented signature variant detects one additional event and two additional controls. The offline full-series reference performs best, as expected for a one-change DGP and access to future observations. It is not evidence for an implementable online advantage.

\subsection{Which component supplies the signal?}

\begin{table}[htp!]
\centering
\caption{Component ablation for the past-calibrated hybrid detector}
\label{tab:component_ablation}
\footnotesize
\setlength{\tabcolsep}{3pt}
\begin{tabular}{p{3.7cm}cccc}
\toprule
\textbf{Variant} & \textbf{Event hit} & \textbf{Delay} & \textbf{Invalid alert} & \textbf{Control alert} \\
\midrule
Shape-only signature & 26\% & 15 & 72\% & 99\% \\
Directional gate only & 90\% & 6 & 10\% & 49\% \\
Cal.-scaled sig., no gate & 33\% & 2 & 67\% & 96\% \\
Cal.-scaled sig. + gate & 94\% & 6 & 6\% & 31\% \\
Cal.+exposure sig. + gate & 96\% & 6 & 4\% & 32\% \\
\bottomrule
\end{tabular}
\end{table}

Table \ref{tab:component_ablation} sharpens the contribution. The mean-drop gate alone recovers 90\% of events but alerts on 49\% of controls. The calibration-scaled signature without the gate recovers only 33\% and alerts on 96\% of controls. Combining them raises recovery to 94\% while reducing control alerts to 31\%. In this DGP, the gate carries most of the recovery signal; the signature threshold acts mainly as a nuisance filter.

The shape-only signature result is poor because independent min--max scaling removes amplitude and level information, while the simulated events are dominated by deterioration in level. This is a useful negative result, not evidence that shape signatures are generally ineffective.

\subsection{Nuisance regimes}

\begin{table}[htp!]
\centering
\caption{False-alert rates by no-event nuisance scenario for selected online methods}
\label{tab:stress_false_alerts}
\scriptsize
\setlength{\tabcolsep}{1.5pt}
\begin{tabular}{lccccccc}
\toprule
\textbf{Nuisance regime} & \textbf{Shape} & \textbf{Cal.} & \textbf{Cal.+I} & \textbf{Feat.} & \textbf{Kernel} & \textbf{GLR} & \textbf{CUSUM} \\
\midrule
Attribution volatility & 100\% & 88\% & 75\% & 83\% & 92\% & 92\% & 100\% \\
Auction pressure & 100\% & 4\% & 8\% & 4\% & 38\% & 38\% & 42\% \\
Creative rotation & 100\% & 0\% & 0\% & 0\% & 0\% & 0\% & 4\% \\
Delivery gaps & 96\% & 21\% & 25\% & 29\% & 50\% & 62\% & 67\% \\
Impression shift & 100\% & 29\% & 25\% & 29\% & 42\% & 50\% & 50\% \\
Seasonality shift & 100\% & 8\% & 17\% & 12\% & 25\% & 25\% & 46\% \\
Stable & 100\% & 4\% & 12\% & 12\% & 29\% & 29\% & 50\% \\
Targeting mix & 100\% & 96\% & 96\% & 92\% & 100\% & 100\% & 100\% \\
\bottomrule
\end{tabular}
\end{table}

Nuisance sensitivity remains material. Attribution volatility and changes in targeting mix frequently trigger every online detector in Table \ref{tab:stress_false_alerts}. In the latter scenario, latent CTR falls because the audience mix changes; the observation is intentionally indistinguishable from a fatigue-like decline without campaign context. The benchmark therefore demonstrates the identification boundary rather than solving it.

\subsection{Train/held-out setting selection}

\begin{table}[htp!]
\centering
\caption{Train-selected thresholds evaluated on held-out synthetic replicates}
\label{tab:tuned_holdout}
\footnotesize
\setlength{\tabcolsep}{3pt}
\resizebox{\textwidth}{!}{%
\begin{tabular}{p{3.8cm}cccc}
\toprule
\textbf{Method / frozen setting} & \textbf{Holdout hit} & \textbf{Control alert} & \textbf{All-series alert} & \textbf{Mean net proxy} \\
\midrule
CUSUM drop (16$\sigma$) & 81\% & 60\% & 71\% & 1262 \\
GLR mean-drop ($k=3.0$) & 86\% & 49\% & 63\% & 1248 \\
Cal.+exposure sig. ($k=3.0$) & 94\% & 25\% & 45\% & 1245 \\
MOSUM mean-drop ($k=2.5$) & 89\% & 52\% & 65\% & 1244 \\
Window features ($k=1.5$) & 89\% & 40\% & 55\% & 1242 \\
Cal.-scaled sig. ($k=3.0$) & 89\% & 26\% & 45\% & 1233 \\
Kernel window ($k=1.0$) & 86\% & 51\% & 64\% & 1212 \\
\bottomrule
\end{tabular}
}
\end{table}

With settings selected by a synthetic click-equivalent objective and then frozen, the signature variants remain competitive but do not dominate. The exposure-augmented variant has the highest held-out event recovery and lowest control-alert rate among the displayed rows, while CUSUM has the highest mean held-out net proxy. The result depends on the arbitrary 150-click review cost and shared scenario families; it cannot establish a generally optimal operating point.

\section{Operational Interpretation and Limitations}
\label{sec:limitations}

\subsection{What the screen can support}

The proposed use is portfolio triage. A periodic batch process could rank creatives for analyst review using the alert, trajectory plot, data-quality checks and benchmark-relative opportunity proxy. Campaign teams would then assess delivery, targeting, placements, audience saturation, creative history and business constraints. The detector should not retire a creative or reallocate spend without that context.

Before operational use, the research detector would need to be implemented in the shared library, tested across the command-line, Python and application interfaces, and evaluated under an explicit alert policy. Threshold selection should reflect review capacity and asymmetric error costs. Portfolio use also creates multiple-testing and dependence problems that are absent from per-series headline rates.

\subsection{Principal limitations}

\begin{enumerate}
    \item \textbf{Synthetic-only evidence.} Scenario magnitudes are designed rather than calibrated from a representative campaign population. One random seed and 24 replicates per scenario do not characterise Monte Carlo uncertainty.
    \item \textbf{Incomplete nuisance library.} The controls cover several plausible confounders but cannot span media-system or reporting changes. The targeting-mix result shows that observed CTR alone does not identify mechanism.
    \item \textbf{No genuine missingness experiment.} All series are complete daily observations. Calendar gaps, irregular timestamps and zero delivery require an explicit path and window convention plus separate evaluation.
    \item \textbf{Heuristic calibration.} The past-only mean-plus-standard-deviation threshold has no false-alarm guarantee under dependent overlapping windows. The smoothing prior, window size, depth, history length and gate are fixed design choices.
    \item \textbf{Comparator scope.} The benchmark includes useful online and offline references but not every strong alternative, such as fully specified state-space models, Bayesian online changepoint detection or exposure-weighted likelihood methods.
    \item \textbf{No causal or financial identification.} The screen and opportunity proxy do not estimate the effect of refreshing a creative, incremental clicks, savings or return on investment.
\end{enumerate}

A decisive next study should use privacy-preserving campaign histories with documented intervention or review dates, prospectively frozen thresholds, pre-specified nuisance strata and an analyst-capacity metric. It should compare the hybrid rule with a strong engineered-feature screen and exposure-aware likelihood or state-space baselines. Genuine missingness and changing delivery should be experimental factors rather than informal claims.

\section{Conclusion}
\label{sec:conclusion}

This paper evaluates path signatures as one component of a creative-fatigue screening workflow. The scientifically accurate contribution is narrower than a signature-based fatigue detector: calibration-scaled signatures measure changes in ordered local trajectory, while a separate mean-drop gate supplies fatigue-oriented deterioration. In the synthetic known-onset benchmark, the hybrid rule recovers most injected events and reduces nuisance alerts relative to the gate alone. A simple engineered window-feature comparator is effectively tied, and several nuisance mechanisms remain difficult.

The result supports continued research, not deployment or causal attribution. Creative fatigue is an old decision problem; path signatures offer a systematic feature representation for testing it. Their value will depend on whether they improve portfolio review decisions after alert burden, campaign context and alternative methods are considered on real, prospectively governed evidence.

\appendix
\section{Signature Qualifications}
\label{app:signature}

The first signature level is total displacement,
\[
S^1(X)=X(1)-X(0).
\]
For two paths \(X^L(u)=(u,z_L(u))\) and \(X^R(u)=(u,z_R(u))\), the time increments are both one. Hence the depth-one contribution to their distance is
\[
\left|(z_L(1)-z_L(0))-(z_R(1)-z_R(0))\right|.
\]
For affine windows this is the absolute difference in endpoint slopes, not the difference in window means. The gate in Equation \ref{eq:gate} is therefore not a signature term in the implemented embedding.

Second-level cross terms encode ordering through iterated interactions. Their antisymmetric part corresponds to signed area; the full second level should not be described simply as area. Higher levels can distinguish some paths with similar endpoints, but truncation loses information. The empirical question is whether the retained terms help the screening objective.

The implementation treats each daily series as an ordinary ordered piecewise-linear path. Missing observations would require a declared convention for calendar time, interpolation or path augmentation. Missing data do not imply that the latent process is discontinuous, and no jump-aware rough-path construction is evaluated.

\section{Supplementary Operating Points and Reproducibility}
\label{app:reproducibility}

\begin{table}[htp!]
\centering
\caption{Operating points for online detectors under matched scoring}
\label{tab:operating_curve}
\footnotesize
\setlength{\tabcolsep}{3pt}
\resizebox{\textwidth}{!}{%
\begin{tabular}{p{4.0cm}cccc}
\toprule
\textbf{Method / setting} & \textbf{Event hit} & \textbf{Median delay} & \textbf{Invalid event alert} & \textbf{No-event false alert} \\
\midrule
Signature shape-only ($k=1.0$) & 8\% & 4 & 92\% & 100\% \\
Signature shape-only ($k=1.5$) & 26\% & 15 & 72\% & 99\% \\
Signature shape-only ($k=2.5$) & 18\% & 22 & 28\% & 56\% \\
Cal.-scaled sig. ($k=1.0$) & 94\% & 6 & 6\% & 35\% \\
Cal.-scaled sig. ($k=1.5$) & 94\% & 6 & 6\% & 31\% \\
Cal.-scaled sig. ($k=2.5$) & 94\% & 6 & 4\% & 26\% \\
Cal.+exposure sig. ($k=1.0$) & 96\% & 6 & 4\% & 36\% \\
Cal.+exposure sig. ($k=1.5$) & 96\% & 6 & 4\% & 32\% \\
Cal.+exposure sig. ($k=2.5$) & 99\% & 6 & 1\% & 24\% \\
Window features ($k=1.0$) & 92\% & 6 & 7\% & 39\% \\
Window features ($k=1.5$) & 94\% & 6 & 4\% & 33\% \\
Window features ($k=2.5$) & 83\% & 5 & 8\% & 24\% \\
Kernel window ($k=1.0$) & 90\% & 10 & 10\% & 49\% \\
Kernel window ($k=1.5$) & 93\% & 11 & 7\% & 47\% \\
Kernel window ($k=2.5$) & 53\% & 13 & 3\% & 32\% \\
GLR mean-drop ($k=1.0$) & 90\% & 6 & 10\% & 49\% \\
GLR mean-drop ($k=1.5$) & 90\% & 6 & 10\% & 49\% \\
GLR mean-drop ($k=2.5$) & 90\% & 6 & 10\% & 49\% \\
MOSUM mean-drop ($k=1.0$) & 29\% & 5 & 71\% & 100\% \\
MOSUM mean-drop ($k=1.5$) & 50\% & 6 & 50\% & 93\% \\
MOSUM mean-drop ($k=2.5$) & 90\% & 6 & 10\% & 48\% \\
CUSUM drop (limit=12$\sigma$) & 62\% & 3 & 38\% & 71\% \\
CUSUM drop (limit=20$\sigma$) & 78\% & 5 & 22\% & 57\% \\
CUSUM drop (limit=24$\sigma$) & 81\% & 5 & 19\% & 52\% \\
\bottomrule
\end{tabular}
}
\end{table}

The operating points show that threshold selection changes alert burden and, for some methods, event timing. Non-monotone hit rates can arise because the score credits only the first alert inside a fixed post-onset window; raising a threshold can suppress an earlier invalid alert.

The benchmark is generated by
\begin{center}
\url{presentations/paper/experiments/injected_changepoint_benchmark.py}.
\end{center}
From the repository root, regenerate all CSV and \LaTeX{} tables with:
\begin{center}
\url{uv run --all-extras python presentations/paper/experiments/injected_changepoint_benchmark.py}
\end{center}
The script records the seed, DGP, detector implementations, scoring rules and output writers. Results are written to \url{presentations/paper/experiments/results/}. Runtime summaries are hardware-dependent and are not used as evidence of deployment readiness.

\end{document}